\documentclass[journal]{IEEEtran}

\ifCLASSINFOpdf
\else
   \usepackage[dvips]{graphicx}
\fi
\usepackage{url}

\usepackage{graphicx}
\usepackage{cite}
\usepackage{amsmath,amsfonts}
\usepackage{algorithmic}
\usepackage{graphicx}
\usepackage{textcomp}
\usepackage{xcolor}
\usepackage{amsthm}
\usepackage{caption}
\usepackage{relsize}
\usepackage{subcaption}
\usepackage{float}
\def\BibTeX{{\rm B\kern-.05em{\sc i\kern-.025em b}\kern-.08em
    T\kern-.1667em\lower.7ex\hbox{E}\kern-.125emX}}
 \usepackage{upgreek}
\usepackage{systeme}
\usepackage{color}
\usepackage{pgfplots}
\usepackage{bbm}
    \DeclareMathOperator{\tr}{tr}
  \newcommand{\figref}[1]{Fig.~\protect\ref{#1}}

                            \newcommand{\pto}{\overset{P}\longrightarrow }

\DeclareMathOperator*{\argmax}{arg\,max}
\DeclareMathOperator*{\argmin}{arg\,min}

\newfont{\bbb}{msbm10 scaled 700}

\newfont{\bb}{msbm10 scaled 1100}

\newcommand{\av}{{\bf a}}
\newcommand{\bv}{{\bf b}}

\newcommand{\ev}{{\bf e}}

\newcommand{\gv}{{\bf g}}

\newcommand{\pv}{{\bf p}}
\newcommand{\qv}{{\bf q}}

\newcommand{\sv}{{\bf s}}

\newcommand{\uv}{{\bf u}}
\newcommand{\wv}{{\bf w}}

\newcommand{\xv}{{\bf x}}
\newcommand{\yv}{{\bf y}}

\newcommand{\Am}{{\bf A}}
\newcommand{\Bm}{{\bf B}}
\newcommand{\Cm}{{\bf C}}

\newcommand{\Hm}{{\bf H}}
\newcommand{\Id}{{\bf I}}

\newcommand{\Rm}{{\bf R}}

\newcommand{\Tm}{{\bf T}}
\newcommand{\Um}{{\bf U}}
\newcommand{\Wm}{{\bf W}}

\newcommand{\Xm}{{\bf X}}
\newcommand{\Ym}{{\bf Y}}

\newcommand{\Zerom}{{\bf 0}}

\newcommand{\Gammam}{\hbox{\boldmath$\Gamma$}}

\newcommand{\Deltam}{\hbox{\boldmath$\Delta$}}

\newcommand{\Phim}{\hbox{\boldmath$\Phi$}}

\renewcommand{\arg}{{\hbox{arg}}}

\newtheorem{theorem}{Theorem}

\newtheorem{remark}{Remark}

\newtheorem{corollary}{Corollary}

\begin{document}
\title{Asymptotic Characterisation of Regularised Zero-Forcing Receiver for Imperfect and Correlated Massive MIMO Systems with Optimal Power Allocation}

%

\author{Ayed M. Alrashdi

 
%
%
\thanks{
The author is with the Department of Electrical Engineering, College of Engineering, University of Ha'il, P.O. Box 2440, Ha'il, 81441, Saudi Arabia (e-mail:
am.alrashdi@uoh.edu.sa).
%
%
%
}
}

\maketitle

\begin{abstract}
In this paper, we present asymptotic high dimensional analysis of the regularised zero-forcing (RZF) receiver in terms of its mean squared error (MSE) and bit error rate (BER) when used for the recovery of binary phase shift keying (BPSK) modulated signals in a massive multiple-input multiple-output (MIMO) communication system. We assume that the channel matrix is spatially correlated and not perfectly known. We use the linear minimum mean squared error (LMMSE) method to estimate the channel matrix.
The asymptotic approximations of the MSE and BER enable us to solve various practical optimisation problems. Under MSE/BER minimisation, we derive 1) the optimal regularisation factor for RZF; 2) the optimal power allocation scheme. Numerical simulations show a close match to the derived asymptotic results even for a few dozens of the problem dimensions.
\end{abstract}
\begin{IEEEkeywords}
Regularisation, spatial correlation, channel estimation, power allocation,  performance analysis, Gaussian min-max theorem.
\end{IEEEkeywords}
\section{Introduction}
\label{sec:intro}
\IEEEPARstart{S}{ince} the early works of \cite{marzetta2010noncooperative,larsson2014massive}, massive multiple-input multiple-output (MIMO) research has been thriving. 

The idea of massive MIMO is to use a very large number of antennas at the base station which offers the desired spatial multiplexing and can reduce the transmitted power \cite{larsson2014massive}. Therefore, it has been considered a promising vital technology to achieve the high spectral/energy efficiencies and high data rates required by the fifth generation (5G) and next wireless communication generations \cite{8187178}.

Channel state information (CSI) plays an important role in attaining the significant benefits of massive MIMO systems, and accurately recovering the transmitted symbols \cite{8187178}. It is well known that perfect knowledge of CSI is an ideal scenario that is impossible to obtain. However, in practice, only imperfect or partial CSI can be acquired through a process called channel estimation or training.
Training refers to the process of sending a known sequence of pilot symbols which can be directly incorporated in the process of estimating the CSI. After this step, the receiver employs the estimated CSI to detect the corresponding transmitted data symbols.

The overall system performance can be improved by optimising the power allocation between the transmitted pilot and data symbols. 
Power optimisation problems in MIMO systems have been proposed based on different performance metrics. In \cite{ballal2019optimum, zhao2017game}, the authors derived a power allocation scheme based on minimising the mean squared error (MSE), while minimising the the bit error rate (BER) and symbol error rate (SER) was considered in \cite{wang2014ber, alrashdi2020optimum, alrashdi2018optimum}.  Training optimisation based on maximising the channel capacity was addressed in \cite{gottumukkala2009, kannu2005capacity, hassibi2003much}.  In addition, the authors in \cite{dao2018pilot, zhu2018uplink, lu2018training} provided power allocation strategies based on maximising the sum rates.
Training optimisation problems are considered in a wide range of systems including
traditional MIMO systems \cite{hassibi2003much}, single-cell massive MIMO systems \cite{cheng2016optimal} and multi-cell multi-user MIMO networks \cite{zhu2018uplink,liu2016pilot,van2018joint}. The list of above references is not inclusive, since power allocation optimisation research has very rich literature. However, we cited the most related works to this paper.


The power allocation in the aforementioned papers was investigated essentially for uncorrelated channel models. However, in practice, wireless communication systems, including massive MIMO systems, are generally spatially correlated \cite{bjornson2017massive}. The power allocation optimisation problem was developed for correlated channels to maximise the sum rates \cite{wagner2012large, muharar2020optimal}, or the spectral efficiency \cite{boukhedimi2018lmmse, cheng2016optimal}. 
To the best of our knowledge, power optimisation problems based on MSE or BER minimisation that involve spatial correlation models in massive MIMO systems are largely unexplored.

In this paper, we propose the use of the regularised zero-forcing (RZF) as a low complexity receiver for a spatially correlated massive MIMO system. We derive novel sharp asymptotic approximations of its MSE and BER performance using binary phase shift keying (BPSK) signaling for simplicity. Then, these approximations are used to derive an optimal power allocation scheme between pilot and data symbols. The main technical tool used in our analysis is the recently developed convex Gaussian min-max Theorem (cGMT) \cite{thrampoulidis2018precise, thrampoulidis2018symbol}.
The cGMT framework has been used to analyse the error performance of various regression and classification problems under independent and identically distributed (i.i.d.) assumption on the entries of the channel matrix \cite{thrampoulidis2018symbol, alrashdi2017precise, alrashdi2019precise, dhifallah2020precise, hayakawa2020asymptotic, kammoun2021precise,deng2019model, salehi2020performance}. For correlated channel matrices, the cGMT was recently used in \cite{alrashdi2020box, alrashdi2020precise} to characterise the performance of the Box-relaxation and the LASSO detectors, respectively. However, these references assume the ideal case of perfect knowledge of the CSI which is impossible to obtain in practice, while this work deals with the more difficult and common in practice scenario of imperfect CSI.
\subsection{Organisation}
The remainder of this paper is structured as follows.
Section \ref{sec:sys model} describes the system model and the considered RZF receiver. The main asymptotic analysis results are presented in Section~\ref{sec:main}. Section~\ref{sec:sim} presents the numerical simulations used to verify the high accuracy of our results.  In addition, Section~\ref{sec:power} illustrates the optimal power allocation scheme derived in this paper.
The paper is then concluded in Section~\ref{sec:concl}.
Finally, the approach of the proof of the main results is given in Appendix~\ref{sec:proof}. 
\subsection{Notations}
Bold face lower case letters (e.g., $\xv$) represent a column vector while $x_i$ is its $i^{th}$ entry and $\| \xv \|$ represents its $\ell_2$-norm. Matrices are denoted by upper case letters such as $\Xm$, with $\Id_n$ being the $n \times n$ identity matrix, while $\mathbf{0}_{m \times n}$ is the all-zeros matrix of size $m \times n$. The $(i,j)$ entry of matrix $\Xm$ is denoted as $[\Xm]_{ij}$. $\tr(\cdot)$, $(\cdot)^T$, and $(\cdot)^{-1}$ are the trace, transpose and inverse operators, respectively.  $\Xm^{1/2}$ represents the square root of matrix $\Xm$ such that $\Xm = \Xm^{1/2}\Xm{^{T/2}}$.We use the standard notations $\mathbb{E}[\cdot],$ and $\mathbb{P}[\cdot]$ to denote the expectation of a random variable, and probability of an event, respectively. The notation $\qv \sim \mathcal{N}(\mathbf{0},\Rm_{\qv})$ is used to denote that the random vector $\qv$ is normally distributed with $\mathbf{0}$ mean and covariance matrix $\Rm_{\qv} = \mathbb{E}[\qv \qv^T]$, where $\mathbf{0}$ represent the all-zeros vector. We write $``\pto "$ to denote convergence in probability as $n \to \infty$.
The notation $f(t)=\mathcal{O}(g(t))$ means $\left|\frac{f(t)}{g(t)}\right|$ is bounded as $t\xrightarrow[]{} \infty$. Finally, $Q(x) = \frac{1}{\sqrt{2 \pi}} \int_{x}^{\infty} e^{-u^2/2} {\rm d}u$ is the $Q$-function associated with the standard normal density.
\section{System Model and Signal Detection}\label{sec:sys model}
\begin{figure}[ht]
\begin{subfigure}{.5\textwidth}
  \centering
\includegraphics[width= 4.9cm]{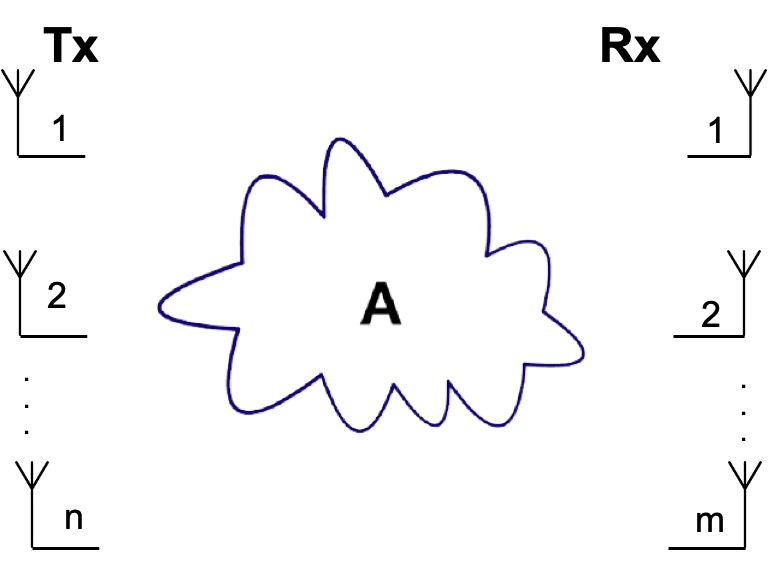}
\caption{\scriptsize{A massive MIMO system.}}
\label{MIMO:fig}
\end{subfigure}
\begin{subfigure}{.5\textwidth}
\centering
\includegraphics[width= 5.9cm]{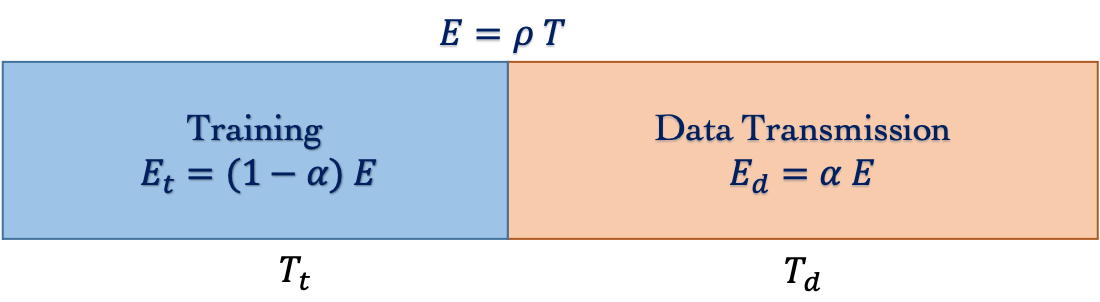}
\caption{\scriptsize{Pilot-based transmission.}}
\label{Training:fig}
\end{subfigure}
\caption{\scriptsize{System model.}}
\label{fig:System model}
\end{figure}

We consider a flat block-fading massive MIMO system with $n$ transmitters (Tx) and $m$ receivers (Rx). The transmission consists of $T$ symbols that occur in a time interval within which the channel is assumed to be constant. A number $T_t$ pilot symbols (for channel estimation) occupy the first part of the transmission interval with power, $\rho_t $. The remaining part is reserved for transmitting $T_d= T - T_t$ data symbols with power, $\rho_d$. It implies from conservation of time and energy that:
\begin{equation}
\label{eq:energy conseve}
\rho_t  T_t + \rho_d T_d = \rho T,
\end{equation}
where $\rho$ is the expected average power.
Alternatively, we have $ \rho_d T_d= \alpha \rho T$, where $\alpha \in (0,1)$ is the ratio of the power allocated to the data, then
$\rho_t  T_t =  (1- \alpha) \rho T$ is the energy of the pilots.
\figref{fig:System model} illustrates the considered system model. 

The received signal model for the \emph{data} transmission phase is given by
\begin{equation}
\yv = \sqrt{\frac{\rho_d}{n}} \Am \xv_0 +\wv,
\end{equation}
where the following model-assumptions hold, except if otherwise stated:
\begin{itemize}
\item The MIMO channel matrix is given by \cite{wagner2012large} 
\begin{align}
\Am = \Rm^{1/2} \Hm.
\end{align}
This matrix model is referred to as the receive-correlated Kronecker model \cite{6172680}. This implies that $\Am$ has $n$ i.i.d.  columns, each with zero mean and covariance matrix $\Rm$.
\item $\Hm \in \mathbb{R}^{m \times n}$ is a random matrix which has i.i.d. standard Gaussian entries (with zero mean and unit variance).
\item $\Rm \in \mathbb{R}^{m \times m}$ is a positive semi-definite Hermitian matrix, satisfying\footnote{We assume this for analytical simplicity.} $\frac{1}{m} \tr(\Rm) = \mathcal{O}(1)$. It captures the spatial correlation between the receive antennas and hence termed the receive-correlation matrix.
\item $\wv \in \mathbb{R}^{m}$ is the noise vector with i.i.d. standard Gaussian entries, i.e., $\wv \sim \mathcal{N}(\mathbf{0},\Id_m)$.
\item $\xv_0 \in \mathbb{R}^{n}$ is the signal to be recovered, which is assumed to be a binary phase shift keying (BPSK) signal, i.e., $\xv_0 \in \{ \pm 1\}^n$.
\end{itemize}
\subsection{Channel Matrix Estimation}
In this paper, we consider the linear minimum mean squared error (LMMSE) estimate $\widehat{\Am}$ of the channel matrix $\Am$, which is given by \cite{kay1993fundamentals}
\begin{align}
\widehat{\Am} = \sqrt{\frac{n}{\rho_t }} \Rm \left(\Rm + \frac{n}{T_t \rho_t } \Id_m \right)^{-1} \Ym_t \Xm_t^T,
\end{align}
where $\Ym_t =  \sqrt{\frac{\rho_t }{n}} \Am \Xm_t + \Wm_t \in \mathbb{R}^{m \times T_t}$ is the received signal corresponding to the \emph{training} phase, $\Xm_t \in \mathbb{R}^{n \times T_t}$ is the matrix of transmitted orthogonal pilot symbols with $T_t \geq n$, and $\Wm_t \in \mathbb{R}^{m \times T_t}$ is an additive white Gaussian noise (AWGN) matrix with $\mathbb{E} [\Wm_t \Wm_t^T]= T_t \Id_m$.

According to \cite{hoydis2013massive, kay1993fundamentals}, the $k^{th}$ column (for all $k\leq n$) of $\widehat{\Am}$ is distributed as
$\mathcal{N}({\bf{0}},\Rm_{\widehat{\Am}})$
with a covariance matrix $\Rm_{\widehat{\Am}}$ that is given by
\begin{align}
\Rm_{\widehat{A}} = \Rm \left(\Rm + \frac{n}{T_t \rho_t } \Id_m \right)^{-1} \Rm.
\end{align}
Note that the pilots energy $T_t \rho_t $ controls the quality of the estimation. In fact, as $T_t \rho_t  \to \infty$, $\widehat{\Am} \to \Am$ which corresponds to the perfect CSI case.

By invoking the orthogonality principle of the LMMSE estimator, it can be shown that the $k^{th}$ column of the \emph{estimation error matrix} $\Deltam := \widehat{\Am} - \Am$ follows the distribution $\mathcal{N}({\bf{0}},\Rm_{\Delta})$ with the following covariance matrix \cite{kay1993fundamentals}:
$$
\mathbf{R}_{\Delta} = \Rm - \Rm_{\small{\widehat{A}}}.
$$
From the orthogonality principle of the LMMSE as well, one can show that $\widehat{\Am}$ and $\Deltam$ are uncorrelated, but both of them follow a Gaussian distribution, hence they are statistically independent.\footnote{{This independence notion is needed in deriving the cGMT results. Since, in \eqref{P,AO}, $\xi(\av,\bv)$ should be independent of $\Am$, then by expressing $\Am$ as $\Am =\widehat{\Am} +\Deltam$, one can ensure that.}}
\subsection{Signal Detection: RZF Receiver}
In this work, we consider the regularised zero-forcing (RZF) receiver that solves the following optimisation
\begin{equation}\label{RZF:eq1}
\widehat{\xv}: = {\rm{arg}} \min_{\xv}  \| \yv - \sqrt{\frac{\rho_d}{n}} \widehat{\Am} \xv \|^2 + \rho_d \lambda \| \xv\|^2,
\end{equation}
where $\lambda \geq 0$ is the regularisation factor. 
For this RZF receiver, $\widehat{\xv}$ admits the following closed-form solution:
\begin{align}
\widehat{\xv} =\left( \widetilde{\Am}^T \widetilde{\Am} + \rho_d \lambda \Id_n \right)^{-1} \widetilde{\Am}^T \yv,
\end{align}
with $\widetilde{\Am} :=\sqrt{\frac{\rho_d}{n}} \widehat{\Am}$.
For this receiver, the detection is performed as follows
\begin{equation}\label{eq:Detection}
{\xv}^*: ={ \rm{sign}}(\widehat{\xv}),
\end{equation}
where ${ \rm{sign}}(\cdot)$ is the sign function which operates element-wise on vector inputs.
\subsection{Figures of Merit}
To evaluate the performance of the RZF receiver, we consider the following performance metrics:\\
{\textit{1) Mean Squared Error (MSE)}}: This measures the performance of the estimation step of the receiver (the first step in \eqref{RZF:eq1}) and is defined as:
\begin{align}
{\rm{MSE}}_n := \frac{1}{n} \| \xv_0 -\widehat{\xv} \|^2.
\end{align}
{\textit{2) Bit Error Rate (BER)}}: This metric is used to evaluate the performance of the second step of the receiver, i.e., the detection step in \eqref{eq:Detection}. It is defined as
\begin{align}
{\rm{BER}}_n := \frac{1}{n} \sum_{i=1}^n {\mathbf{1}}_{\{x_i^* \neq x_{0,i}\}},
\end{align}
where ${\mathbf{1}}_{\{\cdot\}}$ is the indicator function.

In relation to the BER is the \textit{probability of error}, $P_e$, which is defined as the expected value of the BER averaged
over the noise, the channel and the constellation. Formally,
\begin{align}
P_e:= \mathbb{E}[{\rm{BER}}_n] = \frac{1}{n} \sum_{i=1}^n \mathbb{P}\left[\{x_i^* \neq x_{0,i}\}\right].
\end{align}
\section{Main Results}\label{sec:main}
In this section, we provide our main results on the asymptotic characterisation of the RZF receiver in terms of its MSE and BER. 
\subsection{Technical Assumptions}
First, we need to state some technical assumptions that are required for our analytical analysis.\\
\textit{Assumption (1)}: We assume that the problem dimensions $m$ and $n$ are growing large to infinity with a fixed ratio, i.e., $$m \to \infty, n \to \infty, \frac{m}{n} \to \zeta,$$ for some fixed constant $\zeta>0$.\\
\textit{Assumption (2)}: We assume that the normalised coherence time, normalised number of pilot symbols and normalised number data symbols are fixed and given as 
$$
\frac{T}{n} \to \tau \in (1,\infty),
$$
$$
\frac{T_t}{n} \to \tau_t  \in [1,\infty),
$$
and 
\noindent
$$
\frac{T_d}{n} \to \tau_d,
$$
respectively.

Note that under Assumption (2), the covariance matrix of $\widehat{\Am}$ becomes
$$
\Rm_{\widehat{A}} = \Rm \left(\Rm + \frac{1}{\tau_t  \rho_t } \Id_m \right)^{-1} \Rm,
$$
and the time/energy conservation equation in \eqref{eq:energy conseve} becomes 
\begin{align}
\rho_t  \tau_t  + \rho_d \tau_d = \rho \tau.
\end{align}
Finally, define the spectral decomposition of $\Rm_{\widehat{A}}$ as 
\begin{align}
\Rm_{\widehat{A}} = \Um \Gammam \Um^T,
\end{align}
 where $\Um \in \mathbb{R}^{m \times m}$ is an orthonormal matrix, and $\Gammam \in \mathbb{R}^{m \times m}$ is a diagonal matrix with the eigenvalues of $\Rm_{\widehat{A}}$ on its main diagonal.
\subsection{RZF Receiver Performance Characterisation}
\begin{figure}
\begin{center}
\includegraphics[width = 9.1cm]{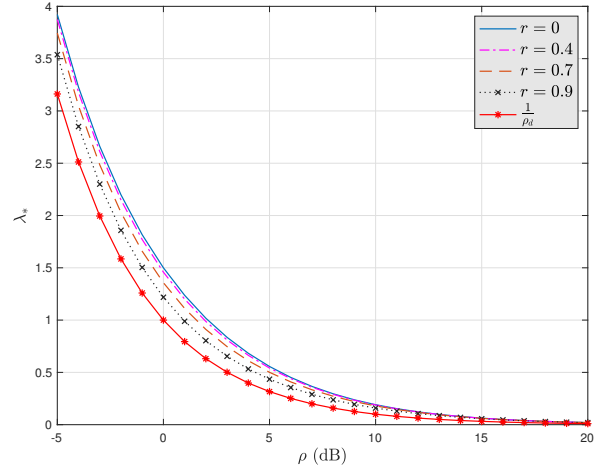}
\end{center}
\caption{\scriptsize {Optimal regulariser $\lambda_*$ v.s. $\rho$ for different correlation coefficients $r$, with $\zeta =1.5, n=500,\alpha=0.5, T=1000, T_t =n$.}}%
\label{fig:reg}
\end{figure}

In this subsection, we precisely characterise the high dimensional performance of the RZF receiver. We begin by stating the MSE analysis as given in the next theorem.
\begin{theorem}[MSE of RZF] \label{Theorm:MSE}\normalfont
 Let $\widehat{\xv}$ be a minimiser of the RZF problem in \eqref{RZF:eq1} for some fixed but unknown BPSK signal $\xv_0$, then for any fixed $\lambda>0, \zeta>0$, and under Assumptions (1) and (2), it holds that
\begin{equation}\label{MSE:RZF}
\left|{\rm{MSE}}_n - \nu_* \right| \pto 0.
\end{equation}
where $\nu_*$ is the unique solution to the following scalar minimax optimisation problem:
\begin{align}
\min_{\nu>0} \max_{\mu>0} \mathcal{F}(\nu,\mu):=&\frac{1}{2n}\sum_{j=1}^{m} \frac{\gamma_{j} \rho_d \nu + \rho_d [\Rm_{\Delta}]_{jj} +1}{\frac{1}{2} + \frac{ \rho_d \gamma_j }{\mu}} \nonumber\\
 &+ \lambda \rho_d( \nu+1) -\frac{ \nu \mu}{2}  -\frac{2 \lambda^2 \rho_d^2 }{\mu},
\end{align}
and $\gamma_j$ is the $j^{th}$ eigenvalue of $\Rm_{\widehat{A}}$.
\end{theorem}
\begin{proof}
The proof of this theorem is given in Appendix~A.
\end{proof}
\begin{remark}\normalfont
From the first order optimality conditions, i.e., 
\begin{align}
\nabla_{(\nu,\mu)} \mathcal{F}(\nu,\mu) = \bf 0,
\end{align}
the solutions $(\nu_*,\mu_*)$ can be easily found as:
\begin{equation}
{\nu}_* =\frac{\frac{1}{n} \sum_{j=1}^{m} \frac{\rho_d \gamma_j ( \rho_d [\Rm_{\Delta}]_{jj} +1 ) }{(\frac{\mu_*}{2}+\rho_d \gamma_j)^2} +\frac{4 \lambda^2 \rho_d^2}{\mu_*^2}}{1-\frac{1}{n} \sum_{j=1}^{m} \frac{\rho_d \gamma_j^2}{(\frac{\mu_*}{2}+\rho_d \gamma_j)^2}},
\end{equation}
and $\mu_*$ is the solution to the following fixed-point equation:
\begin{equation}
\mu_* = \frac{1}{n} \sum_{j=1}^{m} \frac{\mu_* \rho_d \gamma_{j}}{\frac{\mu_*}{2}+\rho_d \gamma_j} + 2 \lambda \rho_d.
\end{equation}
\end{remark}
\begin{remark}\normalfont
For $\Rm = \Id$, (i.e., no correlation, $\gamma_i =1\forall i$,) and perfect CSI ($\Deltam =\mathbf{0}_{m \times n}$), we recover the well-known MSE formula of the Zero-Forcing (ZF) receiver (i.e., when $\lambda=0$):
\begin{align}
\nu_* =\frac{1}{(\zeta-1) \rho_d}.
\end{align}
\end{remark}
Note that the MSE result of Theorem \ref{Theorm:MSE} holds for $\xv_0$ drawn from any distribution with zero mean and unit variance and not necessarily from a BPSK constellation. However, for BPSK signals, the BER of the RZF receiver is given in the next Theorem.
\begin{theorem}[BER of RZF] \label{Theorem:BER}\normalfont
For $\nu>0$, and $\mu>0$, define
\begin{align}
S_{\gamma}(\nu,\mu) :=\frac{1}{n} \sum_{j=1}^{m} \frac{ \rho_d\gamma^2_{j} {\nu} + \rho_d\gamma_j \left(\rho_d [\Rm_{\Delta}]_{jj} +1 \right)}{\big(\frac{1}{2} + \frac{\rho_d \gamma_j }{\mu}\big)^2}.
\end{align}
Then, under the same settings of Theorem~1, it holds that
\begin{equation}\label{BER:RZF}
\left|{\rm{BER}}_n - Q\left(   \sqrt{\frac{4 \lambda^2 \rho_d (1-\nu_*)+ S_{\gamma}(\nu_*,\mu_*)}{\nu_* S_{\gamma}(\nu_*,\mu_*)}} \right) \right| \pto 0.
\end{equation}
\end{theorem}
\begin{proof}
The proof is relegated to Appendix~A.
\end{proof}
\begin{remark}[Probability of error] \normalfont
Recall that $P_e = \mathbb{E}[\rm BER]$, then using \cite[Theorem II.1]{thrampoulidis2018symbol}, we can show that $P_e $ converges to the same asymptotic value as the BER. This means that
\begin{align}
\left|P_e - Q\left(   \sqrt{\frac{4 \lambda^2 \rho_d (1-\nu_*)+ S_{\gamma}(\nu_*,\mu_*)}{\nu_* S_{\gamma}(\nu_*,\mu_*)}} \right) \right| \pto 0.
\end{align}
\end{remark}
\begin{corollary}[Optimal regulariser]\label{opt_reg} \normalfont
The optimal regularisation factor that minimises the MSE or BER is given as
\begin{align}\label{eq:opt_reg}
\lambda_* = \frac{1}{\rho_d} + \frac{1}{m} \tr (\Rm_{\Delta}).
\end{align}
\end{corollary}
\begin{proof}
Note that the MSE expression depends on $\lambda$ through $\nu_*$ only. Hence, the above result can be proven by taking the derivative of $\nu_*$ with respect to $\lambda$. In addition, $\lambda_* $ turned out to be optimal in the BER sense as well. This can be shown by taking the derivative of \eqref{BER:RZF} with respect to $\lambda$.
\end{proof}
\begin{remark}\normalfont
Under perfect CSI ($\Deltam = \Zerom_{m \times n}$), Corollary \ref{opt_reg} simplifies to the well-known formula: $\lambda_* = \frac{1}{\rho_d}$, independent of the correlation matrix $\Rm$, which was previously shown in \cite{wagner2012large, peel2005vector} for other optimality metrics such as maximising the sum rate or SINR, etc.  Here, our optimality metrics are MSE and BER which were not considered for the correlated channel model before.
As mentioned in \cite{peel2005vector}, for large $n$, the RZF receiver is equivalent to the MMSE.
For uncorrelated channels ($\Rm = \Id_m$, and $\Rm_\Delta = \sigma_\Delta^2 \Id_m$), it was proven in \cite{alrashdi2020optimum} that 
\begin{align}
\lambda_* =\frac{1}{\rho_d} + \sigma_\Delta^2,
\end{align}
which is consistent with \eqref{eq:opt_reg} for $\Rm = \Id_m$. 

In \figref{fig:reg}, we use the exponential correlation model for $\Rm$ which is defined as \cite{alfano2004capacity} 
\begin{align}\label{eq:corr1}
\Rm(r) = \bigg[ r^{| i-j|^2} \bigg]_{i,j=1,2,\cdots,m}, r \in [0,1),
\end{align}
 to show the effect of increasing the correlation on the optimal regularisation factor and compare it with the perfect CSI case, i.e., $\lambda_*= \frac{1}{\rho_d}$.
As we can see, for the imperfect CSI scenario, more regularisation is needed due to the channel estimation errors. Furthermore, we observe that as $r$ increases, less regularisation is needed.
\end{remark}
\section{Numerical Results}\label{sec:sim}
To validate our theoretical predictions of the MSE and BER as given by Theorem 1 and Theorem 2, we consider the exponential model given earlier in \eqref{eq:corr1}.
\figref{fig:fig11} shows the MSE/BER curves v.s. the regularisation factor $\lambda$. For the Monte-Carlo (MC) simulations, we used $\zeta =1.5, n=400, r=0.4, \alpha =0.5, T=1000, T_t =n$, and $\rho=10  \ \rm dB$, and the data are averaged over 500 independent Monte-Carlo trials.
We can see that from both figures, there is an optimal value of the regulariser $\lambda_*$ that minimises the MSE/BER. This optimal value is the same for MSE or BER as we can see from the figures.

In addition, we plotted in \figref{fig:mse_SNR}, and \figref{fig:ber_SNR} the MSE/BER performance of the RZF receiver versus the total average power $\rho$ and for different correlation coefficient $r$ values. We used the same parameters values as in the previous experiment. These figures again show the great match between our analytical expressions and the MC simulations.

Finally, in \figref{fig:compare_SNR}, we compare the BER performance of the RZF receiver to the conventional zero-forcing (ZF) receiver (i.e., $\lambda=0$) that is widely used in wireless communications literature. From this figure, it can be seen that the RZF receiver clearly outperforms the ZF. 
\begin{figure}[ht]
\begin{subfigure}{.5\textwidth}
  \centering
%
%
\begin{tikzpicture}[scale=1,font=\small]
    \renewcommand{\axisdefaulttryminticks}{4}
    \tikzstyle{every major grid}+=[style=densely dashed]
    \tikzstyle{every axis y label}+=[yshift=-10pt]
    \tikzstyle{every axis x label}+=[yshift=5pt]
    \tikzstyle{every axis legend}+=[cells={anchor=west},fill=white,
        at={(0.98,0.98)}, anchor=north east, font=\tiny]
        
\begin{axis}	[
width=2.5in,
height=2.0in,
xmin=0,
xmax=1.5,
xlabel={$\lambda$},
ymax=0.59,
ylabel={$\rm MSE$},
      grid=major,
      yticklabels={a,b,$0.1$,$0.2$,$0.3$},
   			]
\addplot [color=red, line width=1.0pt]
  table[row sep=crcr]{%
0.001	0.53408936224628\\
0.011	0.458063914631033\\
0.021	0.407198349444435\\
0.031	0.371061377685961\\
0.041	0.344348529953736\\
0.051	0.324051349683903\\
0.061	0.308328033013901\\
0.071	0.295981571303633\\
0.081	0.286199480798113\\
0.091	0.278404332154927\\
0.101	0.272174062461184\\
0.111	0.267238185154928\\
0.121	0.263317811585335\\
0.131	0.260257408092062\\
0.141	0.257863868137966\\
0.151	0.256092879769065\\
0.161	0.254824934325813\\
0.171	0.25396998628941\\
0.181	0.253469599510137\\
0.191	0.253308814908556\\
0.201	0.253407017754599\\
0.211	0.253737151564417\\
0.221	0.254270955099922\\
0.231	0.254982507268772\\
0.241	0.255849575479676\\
0.251	0.256852999623111\\
0.261	0.257976195077453\\
0.271	0.259190568506776\\
0.281	0.260532722634511\\
0.291	0.261932133215421\\
0.301	0.26340669852052\\
0.311	0.264943129593129\\
0.321	0.266535677729347\\
0.331	0.268177695936395\\
0.341	0.269863274829158\\
0.351	0.271587149480089\\
0.361	0.273344619706327\\
0.371	0.275131481575024\\
0.381	0.276958599187024\\
0.391	0.278795166597778\\
0.401	0.280650759016123\\
0.411	0.282522664726035\\
0.421	0.284408438568658\\
0.431	0.286305873176469\\
0.441	0.288212973733835\\
0.451	0.290127935747764\\
0.461	0.292049125470042\\
0.471	0.293975062558003\\
0.481	0.295904404746228\\
0.491	0.297835934240955\\
0.501	0.299768545584935\\
0.511	0.30170123488669\\
0.521	0.303633090222152\\
0.531	0.305563282963038\\
0.541	0.307491060116417\\
0.551	0.30941573740066\\
0.561	0.311336693024971\\
0.571	0.313253362099412\\
0.581	0.315165231608114\\
0.591	0.317071835817739\\
0.601	0.318972752270066\\
0.611	0.320867597937606\\
0.621	0.322756025965092\\
0.631	0.324637722547972\\
0.641	0.326512404186431\\
0.651	0.328379815156336\\
0.661	0.330239725208549\\
0.671	0.332091927509262\\
0.681	0.333936236669801\\
0.691	0.335772487106194\\
0.701	0.337600531306568\\
0.711	0.339420238573125\\
0.721	0.341231493288633\\
0.731	0.343034194380148\\
0.741	0.344828253518272\\
0.751	0.346613594454275\\
0.761	0.348390151969942\\
0.771	0.350156382505859\\
0.781	0.35191373956489\\
0.791	0.353662241210521\\
0.801	0.355401854658087\\
0.811	0.357132554725272\\
0.821	0.358854323234676\\
0.831	0.360567148464864\\
0.841	0.362271024644274\\
0.851	0.363965951482643\\
0.861	0.365651933738784\\
0.871	0.367328980820358\\
0.881	0.368997106414579\\
0.891	0.370656328145241\\
0.901	0.372306667256182\\
0.911	0.373948148318715\\
0.921	0.375580798956392\\
0.931	0.377204649597536\\
0.941	0.378819733236677\\
0.951	0.380426085222735\\
0.961	0.382064960921135\\
0.971	0.383654049705949\\
0.981	0.385234524398959\\
0.991	0.386806427900305\\
1.001	0.388369804562187\\
1.011	0.389924700095382\\
1.021	0.391471161417574\\
1.031	0.393009236537832\\
1.041	0.394538974445247\\
1.051	0.396060425005248\\
1.061	0.397573638863674\\
1.071	0.39907866735179\\
1.081	0.400575562431887\\
1.091	0.40206437654724\\
1.101	0.403545162653765\\
1.111	0.405017974064041\\
1.121	0.406482864432043\\
1.131	0.407939887681044\\
1.141	0.409389097952378\\
1.151	0.410830549555328\\
1.161	0.41226429692107\\
1.171	0.413690394559751\\
1.181	0.415108897020788\\
1.191	0.416519858857907\\
1.201	0.417923334582479\\
1.211	0.419319378683598\\
1.221	0.420708045450101\\
1.231	0.422089389193489\\
1.241	0.423463463988095\\
1.251	0.42483032376247\\
1.261	0.426190022256622\\
1.271	0.427542612995185\\
1.281	0.428888149283409\\
1.291	0.430226684168602\\
1.301	0.431558270458478\\
1.311	0.432882960666352\\
1.321	0.434200807027979\\
1.331	0.435511861476543\\
1.341	0.436816175640242\\
1.351	0.438113800823618\\
1.361	0.439404788003132\\
1.371	0.440689187819132\\
1.381	0.441967050567591\\
1.391	0.44323842619484\\
1.401	0.444503364290787\\
1.411	0.445761914083621\\
1.421	0.447014124435937\\
1.431	0.448260043839938\\
1.441	0.449499720414613\\
1.451	0.450733201902484\\
1.461	0.451955165451959\\
1.471	0.453176510279635\\
1.481	0.454391798031929\\
1.491	0.455600393286689\\
1.501	0.456802862279553\\
1.511	0.457999430483819\\
1.521	0.459190142505709\\
1.531	0.460375042581308\\
1.541	0.461554174575992\\
1.551	0.462727581984288\\
1.561	0.46389530792987\\
1.571	0.465057395165911\\
1.581	0.466213886075642\\
1.591	0.467364822672981\\
1.601	0.468510246603856\\
1.611	0.469650199146432\\
1.621	0.470784721212854\\
1.631	0.47191385334901\\
1.641	0.473037635743725\\
1.651	0.474156108218805\\
1.661	0.475269310238088\\
1.671	0.476377280908122\\
1.681	0.477480058979948\\
1.691	0.478577682851458\\
1.701	0.479670190567945\\
1.711	0.480757619827902\\
1.721	0.481840007981848\\
1.731	0.482917392036465\\
1.741	0.483989808656771\\
1.751	0.485057294168584\\
1.761	0.486119884561517\\
1.771	0.487177615490836\\
1.781	0.488230522280826\\
1.791	0.489278639927124\\
1.801	0.490322003099503\\
1.811	0.491360646144808\\
1.821	0.492394603089465\\
1.831	0.493423907642514\\
1.841	0.494448593198047\\
1.851	0.495468692838421\\
1.861	0.496484239338652\\
1.871	0.497495265160897\\
1.881	0.498501802476778\\
1.891	0.49955101620273\\
1.901	0.500548718848863\\
1.911	0.501542027431172\\
1.921	0.502530972931077\\
1.931	0.503515586040766\\
1.941	0.504495897166225\\
1.951	0.505471936429883\\
1.961	0.506443733673732\\
1.971	0.507411318461779\\
1.981	0.508374720083031\\
1.991	0.509333967554301\\
2.001	0.510289089622942\\
};
\addlegendentry{Analytical}
\addplot [only marks, line width =1pt, mark size=1.7pt, mark=o, mark options={solid, blue}]
  table[row sep=crcr]{%
0.001	0.535173256772911\\
0.051	0.324756659152196\\
0.151	0.255068553994933\\
0.251	0.258893002947498\\
0.351	0.27148768853525\\
0.451	0.291019352144831\\
0.551	0.308288910655009\\
0.651	0.327046963200367\\
0.751	0.349724720450001\\
0.851	0.365791716063008\\
0.951	0.381554695410138\\
1.051	0.396642669965105\\
1.151	0.410773286107559\\
1.251	0.424535732086705\\
1.351	0.438151245453989\\
1.451	0.452275685636966\\
1.551	0.463499877145545\\
1.651	0.472759582857453\\
1.751	0.486382393202013\\
1.851	0.497952635707522\\
1.951	0.506207900421656\\
2.001	0.510724865615674\\
};
\addlegendentry{MC Simulation}

\end{axis}

\end{tikzpicture} \vskip-4mm
\centering
\caption{\scriptsize {MSE performance.}}%
  \label{fig:sub-first}
\end{subfigure}
\begin{subfigure}{.5\textwidth}
  \centering
\begin{tikzpicture}[scale=1,font=\small]
    \renewcommand{\axisdefaulttryminticks}{4}
    \tikzstyle{every major grid}+=[style=densely dashed]
    \tikzstyle{every axis y label}+=[yshift=-10pt]
    \tikzstyle{every axis x label}+=[yshift=5pt]
    \tikzstyle{every axis legend}+=[cells={anchor=west},fill=white,
        anchor=north east, font=\tiny ]
\begin{semilogyaxis}[
width=2.5in,
height=2.0in,
xmin=0,
xmax=1.5,
xlabel={$\lambda$},
 grid=major,
 scaled ticks=true,
ylabel={$\rm BER$},
yticklabels={0,$10^{-2}$},
]

\addplot [color=red, line width=1.0pt]
  table[row sep=crcr]{%
0.001	0.0862034517656051\\
0.011	0.0753862754524743\\
0.021	0.0677859155705575\\
0.031	0.0622097750134517\\
0.041	0.0579946543387545\\
0.051	0.054739219008373\\
0.061	0.0521860485500071\\
0.071	0.0501614916526118\\
0.081	0.0485446951614432\\
0.091	0.047246890490303\\
0.101	0.046201387278535\\
0.111	0.0453738618634197\\
0.121	0.0447111436928644\\
0.131	0.0441939970837092\\
0.141	0.0437800725321755\\
0.151	0.0434777375030722\\
0.161	0.0432614620645306\\
0.171	0.0431128717480227\\
0.181	0.0430229550122027\\
0.191	0.042998120357116\\
0.201	0.0430157835647759\\
0.211	0.0430735800443455\\
0.221	0.0431673474905453\\
0.231	0.0432927508775907\\
0.241	0.0434460590341076\\
0.251	0.0436240282066286\\
0.261	0.0438238623194536\\
0.271	0.0440368090319476\\
0.281	0.0442825297749541\\
0.291	0.0445326865248346\\
0.301	0.0447979003862478\\
0.311	0.0450744846937391\\
0.321	0.0453619327324576\\
0.331	0.0456590816902762\\
0.341	0.0459648880876519\\
0.351	0.0462784133301512\\
0.361	0.0465988127023619\\
0.371	0.0469253250804084\\
0.381	0.0472641971322173\\
0.391	0.0476018244378207\\
0.401	0.0479436180703389\\
0.411	0.0482890646110551\\
0.421	0.048637706777865\\
0.431	0.0489891280467829\\
0.441	0.0493429486128669\\
0.451	0.0496988293895478\\
0.461	0.0500564477770808\\
0.471	0.0504155132163198\\
0.481	0.050775772466241\\
0.491	0.0511369303884257\\
0.501	0.0514988082765664\\
0.511	0.051861169763906\\
0.521	0.052223894322124\\
0.531	0.0525867564830492\\
0.541	0.0529496023389201\\
0.551	0.0533122870545059\\
0.561	0.053674677370635\\
0.571	0.0540366534824322\\
0.581	0.0543981246177868\\
0.591	0.0547589133799917\\
0.601	0.05511902602779\\
0.611	0.0554782716935262\\
0.621	0.0558366273036787\\
0.631	0.0561940266108332\\
0.641	0.0565503716307565\\
0.651	0.0569056008415767\\
0.661	0.0572597358809532\\
0.671	0.0576125913003683\\
0.681	0.0579641894810394\\
0.691	0.0583144756454584\\
0.701	0.0586634199684375\\
0.711	0.0590109546248853\\
0.721	0.0593571314655048\\
0.731	0.0597018356551906\\
0.741	0.060045013405697\\
0.751	0.060386714551178\\
0.761	0.060726889096519\\
0.771	0.0610646932527089\\
0.781	0.06140099929605\\
0.791	0.0617356890722286\\
0.801	0.062068812909416\\
0.811	0.0624003566463745\\
0.821	0.0627303065338422\\
0.831	0.0630586610820301\\
0.841	0.0633854022438738\\
0.851	0.0637105218460546\\
0.861	0.0640340130067246\\
0.871	0.0643558723455646\\
0.881	0.0646760905942821\\
0.891	0.0649946459715965\\
0.901	0.0653115825864007\\
0.911	0.0656268715932105\\
0.921	0.0659405343383367\\
0.931	0.0662525172702897\\
0.941	0.0665628577489865\\
0.951	0.0668715607780924\\
0.961	0.0672026304464621\\
0.971	0.0675081746532271\\
0.981	0.0678120933214538\\
0.991	0.0681143351166185\\
1.001	0.068414964116539\\
1.011	0.0687139666543576\\
1.021	0.0690113397462926\\
1.031	0.0693071060288887\\
1.041	0.0696012642437875\\
1.051	0.0698938204744066\\
1.061	0.0701847810404679\\
1.071	0.0704741350128225\\
1.081	0.0707619630659189\\
1.091	0.0710481493662875\\
1.101	0.0713328124958258\\
1.111	0.0716158984903767\\
1.121	0.0718974200162448\\
1.131	0.0721774086378991\\
1.141	0.0724558612962928\\
1.151	0.0727327862396388\\
1.161	0.0730081919033729\\
1.171	0.0732820868217923\\
1.181	0.0735544825623536\\
1.191	0.0738253515005038\\
1.201	0.0740948112658868\\
1.211	0.074362725552378\\
1.221	0.0746292060085262\\
1.231	0.0748942164795671\\
1.241	0.0751577815812561\\
1.251	0.0754199105654893\\
1.261	0.0756805988633481\\
1.271	0.0759398979682389\\
1.281	0.0761977753558097\\
1.291	0.0764541993003686\\
1.301	0.0767092708918567\\
1.311	0.0769629494174475\\
1.321	0.0772152529783399\\
1.331	0.0774662450972424\\
1.341	0.077715826282571\\
1.351	0.0779640602465386\\
1.361	0.0782109562021878\\
1.371	0.0784565247264854\\
1.381	0.0787007796033375\\
1.391	0.0789437242060942\\
1.401	0.0791853676643073\\
1.411	0.079425719077516\\
1.421	0.0796647875126325\\
1.431	0.0799025793078323\\
1.441	0.0801391086728508\\
1.451	0.080374382051767\\
1.461	0.0806049135986212\\
1.471	0.0808377680269584\\
1.481	0.0810693914683185\\
1.491	0.0812993466111026\\
};
\addlegendentry{Analytical}
\addplot [only marks, line width =1pt, mark size=1.7pt, mark=o, mark options={solid, blue}]
  table[row sep=crcr]{%
0.001	0.0858816666666666\\
0.051	0.0542083333333333\\
0.101	0.0463449999999999\\
0.151	0.0435633333333332\\
0.201	0.0429449999999998\\
0.251	0.0440999999999998\\
0.301	0.0452516666666665\\
0.351	0.0472149999999999\\
0.401	0.048475\\
0.451	0.0497983333333332\\
0.501	0.051245\\
0.551	0.0532616666666667\\
0.601	0.055475\\
0.651	0.0567383333333333\\
0.701	0.0590400000000001\\
0.751	0.0605766666666667\\
0.801	0.0622416666666667\\
0.851	0.064215\\
0.901	0.0660116666666667\\
0.951	0.0673416666666667\\
1.001	0.069355\\
1.051	0.0696233333333332\\
1.101	0.07137\\
1.151	0.0726433333333332\\
1.201	0.0745033333333333\\
1.251	0.0755216666666667\\
1.301	0.0774833333333332\\
1.351	0.0791233333333333\\
1.401	0.0795133333333333\\
1.451	0.0806283333333332\\
};
\addlegendentry{MC Simulation}
\end{semilogyaxis}
\end{tikzpicture}%
  \label{fig:sub-second}
  \centering
  \caption{\scriptsize {BER Performance.}}
\end{subfigure}
\caption{\scriptsize {Performance of RZF receiver v.s. the regulariser $\lambda$. We used $\zeta =1.5, n=400, r=0.4, T_t =n, T=1000, \alpha =0.5, \rho=10 \rm \ dB$.}}
\label{fig:fig11}
\end{figure}
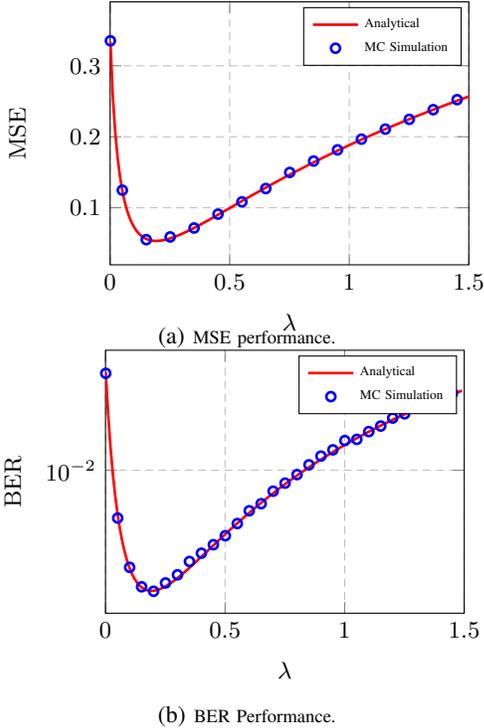
\begin{figure}
\begin{center}
\includegraphics[width = 9.1cm]{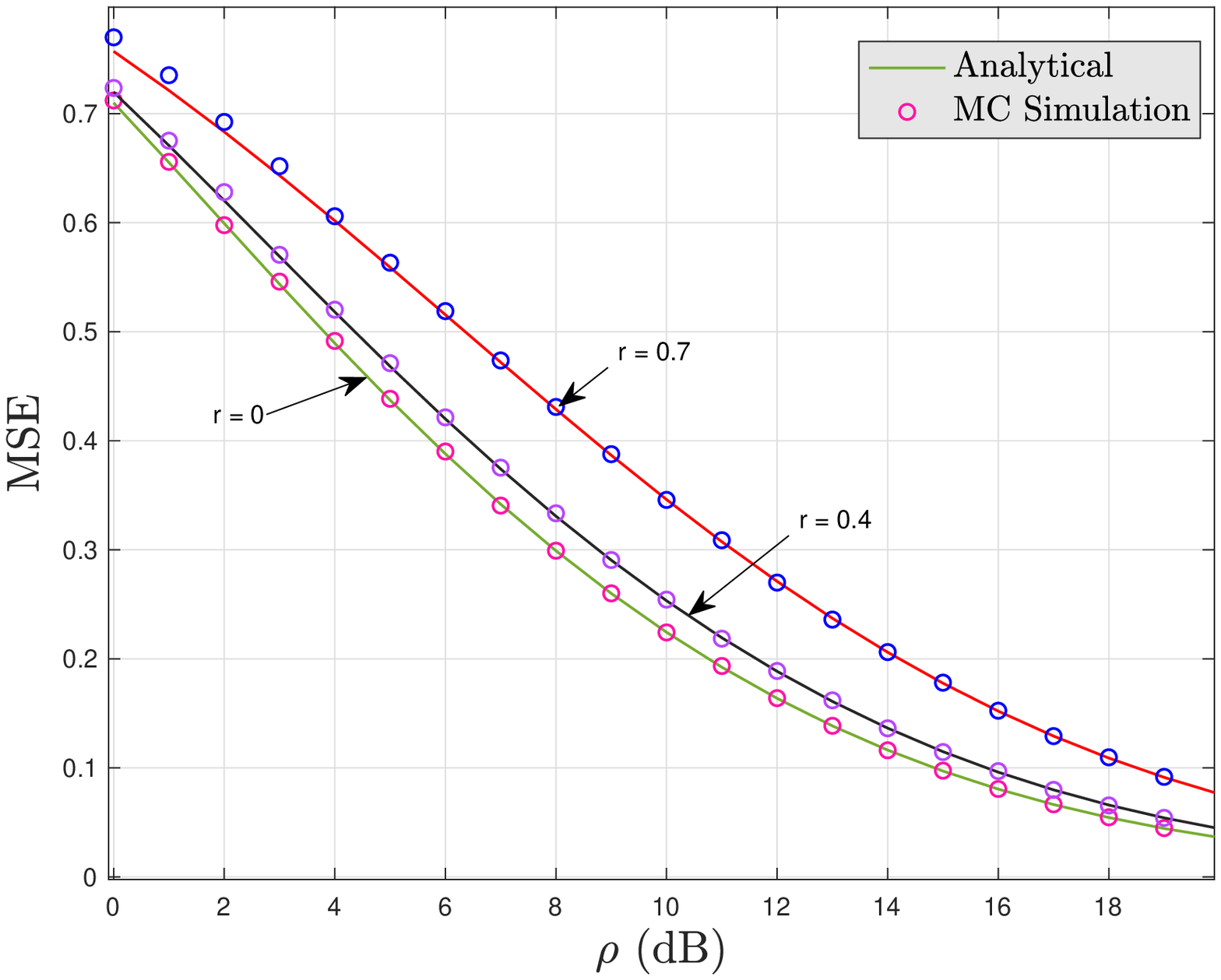}
\end{center}
\caption{\scriptsize{$\rm MSE$ performance of RZF receiver v.s.  $\rho$ for different correlation coefficients $r$, with $\zeta =1.5, n=500,\alpha=0.5, T=1000, T_t =n$.}}%
\label{fig:mse_SNR}
\end{figure}
\begin{figure}
\begin{center}
\includegraphics[width = 9.1cm]{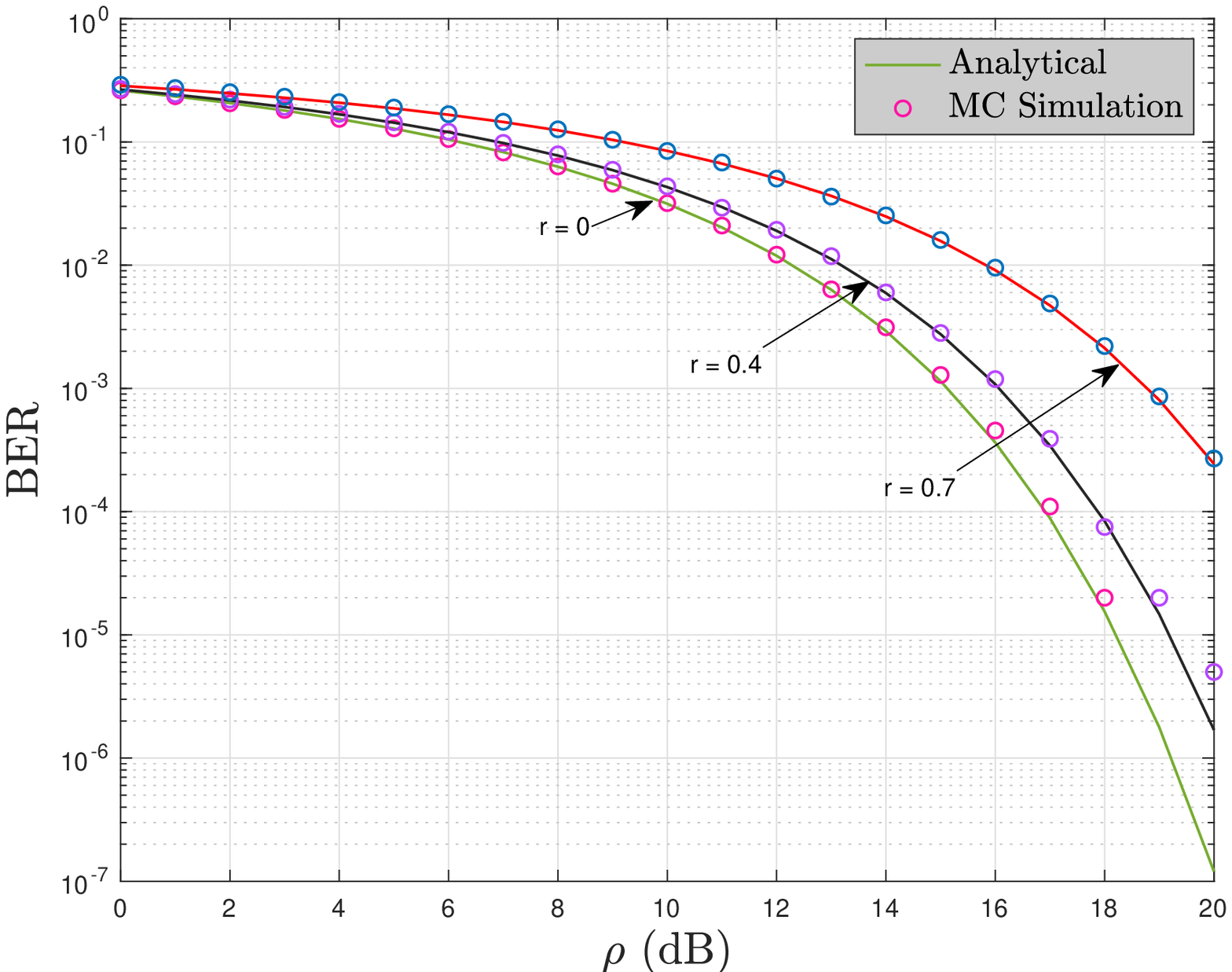}
\end{center}
\caption{\scriptsize{$\rm BER$ performance of RZF receiver v.s.  $\rho$ for different correlation coefficients $r$, with $\zeta =1.5, n=500,\alpha=0.5, T=1000, T_t =n$.}}%
\label{fig:ber_SNR}
\end{figure}
\begin{figure}
\begin{center}
\includegraphics[width = 9.1cm]{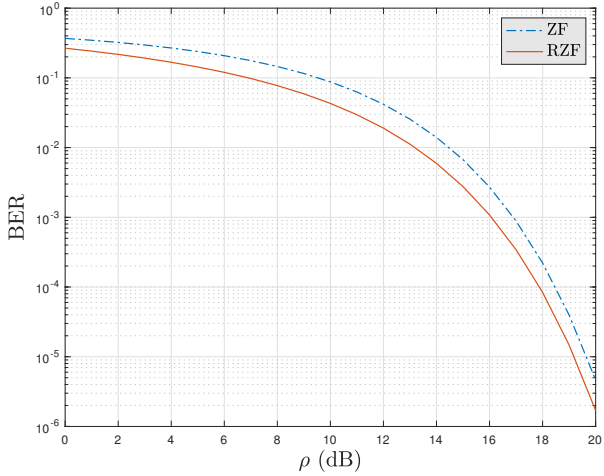}
\end{center}
\caption{\scriptsize{Comparison between BER performance of ZF and RZF receivers. We used the same parameter values as in \figref{fig:ber_SNR}, but with $r =0.4$.}}%
\label{fig:compare_SNR}
\end{figure}
\section{Power Allocation Optimisation}\label{sec:power}
\begin{figure}[htt]
\begin{center}
\includegraphics[width = 9.1cm]{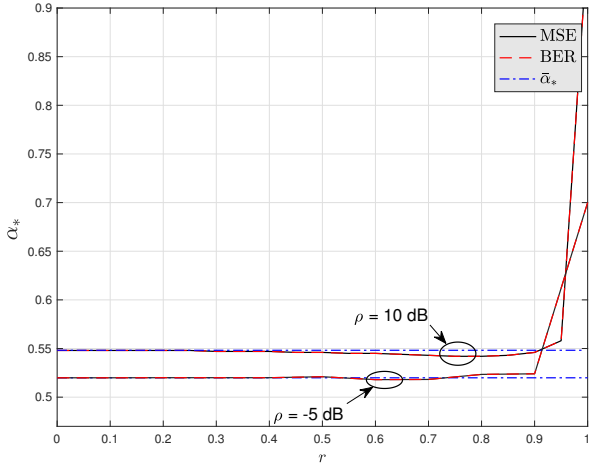}
\end{center}
\caption{\scriptsize {Optimal data power ratio $\alpha_*$ v.s. correlation coefficient $r$, with $\zeta =1.5, n=400,T=1000, T_t =n$.}}%
\label{fig:power}
\end{figure}

%
%
%
In this section, we will use the previous asymptotic approximations of the MSE and BER to find the optimum power allocation between pilot and data symbols to asymptotically minimise the MSE or BER. For fixed $\tau_t $ and $\tau$, the power allocation optimisation problem can be caste as
\begin{align*}
&\min_{\rho_t , \rho_d} \rm{MSE} \\
& \text{subject to:} \ \rho_t  \tau_t  + \rho_d (\tau-\tau_t ) = \rho \tau, \\
& \rho_t  = (1-\alpha) \rho \tau, \rho_d  = \alpha \rho \tau, 0<\alpha<1.
\end{align*}
It can be shown that the above optimisation problem boils down to only optimising the data power ratio $\alpha$, i.e.,
\begin{align}
\alpha_*^{\rm{MSE}} = \argmin_{0<\alpha<1} \mathrm{MSE}(\lambda_*),
\end{align}
where $\mathrm {MSE}(\lambda_*)$ is the asymptotic MSE expression in \eqref{MSE:RZF} while using the optimal value of the regulariser $\lambda_*$ there. Similarly, we have 
\begin{align}
\alpha_*^{\rm{BER}} = \argmin_{0<\alpha<1} \mathrm{BER}(\lambda_*),
\end{align}
where where $\mathrm {BER}(\lambda_*)$ is the asymptotic BER expression in \eqref{BER:RZF}, but with optimal $\lambda_*$.
However, based on \eqref{BER:RZF}, since minimising the $Q$-function amounts to maximising its argument, we have
\begin{align}
\alpha_*^{\rm{BER}} = \argmax_{0<\alpha<1} \mu_*.
\end{align}
For this RZF receiver, finding $\alpha_*^{\rm{MSE}}$ or $\alpha_*^{\rm{BER}}$ in a closed form seems to be a difficult task, but by using a bisection method we can numerically find the optimal power allocation as shown in \figref{fig:power} for different values of the correlation coefficient $r$. In \cite{alrashdi2020optimum}, for the uncorrelated channel $\Rm =\Id_m$, it has been shown that $ \bar\alpha_*^{\rm{MSE}}=\bar\alpha_*^{\rm{BER}} = \bar\alpha_*$, where $\bar\alpha_*$ has the following closed-form expression (see \cite[eq. (36)]{alrashdi2020optimum}):
\begin{equation}\label{optimal_power}
\bar\alpha_* =
\begin{cases}
\vartheta  - \sqrt{\vartheta (\vartheta  -1)},  & \text{if $\tau_{d} > 1$,} \\
\frac{1}{2}, & \text{if $\tau_{d} =1$,} \\
\vartheta  + \sqrt{\vartheta (\vartheta  -1)} & \text{if $\tau_d< 1$,}
\end{cases}
\end{equation}
where $\vartheta = \frac{1 + \rho \tau}{\rho \tau (1 - \frac{1}{\tau_d})}$.
From \figref{fig:power}, we can see that even for the correlated case, we still have that $\alpha_*^{\rm{MSE}}=\alpha_*^{\rm{BER}}$ which indicates that optimising the MSE is equivalent to optimising the BER asymptotically. Furthermore, from this figure we can see that $\bar\alpha_*$ is a quite good approximation of $\alpha_*$ for $r \in [0,0.9]$. This suggests that we can use the optimal $\bar \alpha_*$ from the uncorrelated channel model for the correlated channel case with negligible effect on the performance. Similar observations were found in \cite{muharar2020optimal}.
\section{Conclusion}\label{sec:concl}
This work sharply characterises the asymptotic behaviour of the RZF receiver under the presence of correlation and uncertainties (in the form of estimation errors) in the channel matrix. Particularly, we derived asymptotic expressions of the MSE and BER of the RZF. We then considered a concrete application of our theoretical results to a BPSK modulated massive MIMO wireless communication system, and optimise its performance by optimally allocating power between pilot and data symbols. The results also enabled us to set the regularisation factor in an optimal way which was shown to further improves the performance. Numerical results showed great agreement to the derived theoretical expressions even when the dimensions are not very large.

Possible future extensions of this work include: studying more involved modulation schemes (such as PAM, QAM, and PSK), and analysing advanced receivers such as the RZF with a box-constraint. Another interesting future work is to consider the performance of double-sided correlated massive MIMO systems and study their optimal power allocation.
\section*{Acknowledgment}
The work of Ayed M. Alrashdi is supported by the University of Ha'il, Saudi Arabia.
\begin{appendices}
\section{Approach of the Proof}\label{sec:proof}
In this section, we prove the main results of the \textit{RZF receiver}.  We first introduce the main tool used in the analysis, i.e., the cGMT.
\subsection{cGMT Framework}
The proof is based on the cGMT framework \cite{thrampoulidis2018precise}. Here, we recall the statement of the theorem, and we refer the reader to \cite{thrampoulidis2018precise, thrampoulidis2018symbol} for the complete technical details.
Consider the following two min-max problems, which we refer to, respectively, as the Primal Optimisation (PO) and Auxiliary Optimisation (AO):
\begin{subequations}
\begin{align}\label{P,AO}
& \text{(PO)} \quad \Psi^{(n)}(\Cm) := \underset{\av \in \mathcal{K}_{a}}{\operatorname{\min}}  \ \underset{\bv \in \mathcal{K}_{b}}{\operatorname{\max}} \ \bv^{T} \Cm \av + \xi( \av, \bv), \\
& \text{(AO)} \ \psi^{(n)}(\gv_1, \gv_2)\! := \!\underset{\av \in \mathcal{K}_{a}}{\operatorname{\min}}  \underset{\bv \in \mathcal{K}_{b}}{\operatorname{\max}} \| \av \| \gv_1^{T} \bv \!- \!\| \bv \| \gv_2^{T} \av \!+\! \xi( \av, \bv), \label{AA2}
\end{align}
\end{subequations}
where $\Cm \in \mathbb{R}^{\tilde m \times \tilde n}, \gv_1 \in \mathbb{R}^{\tilde m}, \gv_2 \in \mathbb{R}^{\tilde n}, \mathcal{K}_a \subset \mathbb{R}^{\tilde n}, \mathcal{K}_b \subset \mathbb{R}^{\tilde m}$ and $\xi: \mathbb{R}^{\tilde n} \times \mathbb{R}^{\tilde m} \mapsto \mathbb{R}$. Moreover, the function $\xi$ is assumed to be independent of the matrix $\Cm$. Denote by $\av_{\Psi} := \av_{\Psi}(\Cm) $, and $\av_{\psi} := \av_{\psi}( \gv_1, \gv_2)$ any optimal minimisers of (\ref{P,AO}) and (\ref{AA2}), respectively. Further let $\mathcal{K}_a, \mathcal{K}_b$ be convex and compact sets, $\xi(\av,\bv)$ is convex-concave continuous on $\mathcal{K}_a \times \mathcal{K}_b$, and $\Cm, \gv_1$ and $\gv_2 $ all have i.i.d. standard normal entries.
Then, the cGMT framework relates the optimiser $\av_{\Psi}$ of the PO with the optimal value of the AO as summarised in the following theorem.
\begin{theorem}[cGMT, \cite{thrampoulidis2018precise}]
Let $\mathcal{K}$ be any arbitrary open subset of $\mathcal{K}_a $, and $\mathcal{K}^c = \mathcal{K}_a \setminus\mathcal{K}$. Denote $\psi_{\mathcal{K}^c}^{(n)}(\gv_1,\gv_2)$ the optimal cost of the optimisation in (\ref{AA2}), when the minimisation over $\av$ is constrained over $\av \in \mathcal{K}^c$. Suppose that there exist constants $\beta$ and $\delta >0$ such that in the limit as $\tilde n \rightarrow + \infty$, it holds with probability approaching one: $(i) \  \psi^{(n)}(\gv_1,\gv_2) \leq \beta +\delta$, and, $(ii) \ \psi_{\mathcal{K}^c}^{(n)}(\gv_1,\gv_2) \geq \beta + 2\delta$.
Then, 
\begin{align}
\lim_{\tilde n \rightarrow \infty} \mathbb{P}[\av_{\Psi} \in \mathcal{K}] = 1.
\end{align}
\end{theorem}
After introducing the cGMT, we are now in a position to outline the proof of Theorem~\ref{Theorm:MSE} and Theorem~\ref{Theorem:BER}. The steps of the proof are given in the next subsections. 
\subsection{Deriving the Minimax Optimisation}
For convenience, we consider the \textit{error vector} $\ev := \xv- \xv_0 $, 
then the problem in (\ref{RZF:eq1}) can be reformulated as
\begin{equation}\label{eq:w}
\widehat{\ev} = \arg\min_{\ev } \| \sqrt{\frac{\rho_d}{n}}\widehat{\Am} \ev +\sqrt{\frac{\rho_d}{n}} \Deltam \xv_0-\wv \|^2 +\lambda \rho_d \|\ev +\xv_0 \|^2.
\end{equation}
Without loss of generality, we assume that 
\begin{align}
\xv_0 =\mathbf{1}_n= [1,1,\cdots,1]^T. 
\end{align}
Then, 
\begin{align}
{\rm BER} = \frac{1}{n} \sum_{i=1}^n{\mathbf{1}}_{\{\widehat{e}_i \leq -1\}}.
\end{align}
Next, we note that $\widehat{\Am}$ can be written as 
$
\widehat{\Am} = \Rm_{\small{\widehat{A}}}^{1/2}\Bm,
$
with $\Bm$ being a Gaussian matrix with i.i.d. standard entries (0-mean and unit-variance) and $\Rm_{\widehat{\Am}}$ is the covariance matrix of $\widehat{\Am}$ as defined before. Thus, we have
\begin{align}\label{eq:w2}
\widehat{\ev} = \arg\min_{\ev }& \left\| \sqrt{\frac{\rho_d}{n}} \Rm_{\small{\widehat{A}}}^{1/2}\Bm \ev +\sqrt{\frac{\rho_d}{n}} \Deltam \xv_0-\wv \right\|^2 \nonumber\\
&+\lambda \rho_d \|\ev +\xv_0 \|^2,
\end{align}
Since the Gaussian distribution is invariant under orthogonal transformations, and recalling that the spectral decomposition of $\Rm_{\widehat{\Am}}$ is $\Rm_{\widehat{\Am}} = \Um \Gammam \Um^T$, we have
\begin{align}\label{eq:Gamma}
\widehat{\ev} = \arg\min_{\ev }& \left\| \sqrt{\frac{\rho_d}{n}} \Gammam^{1/2}\Bm \ev +\sqrt{\frac{\rho_d}{n}} \Deltam \xv_0-\wv \right\|^2 
\nonumber\\
&+\lambda \rho_d \|\xv_0+\ev\|^2,
\end{align}
with abuse of notation for $\Bm$.\footnote{We reused $\Bm$ to denote another standard Gaussian matrix.}
The loss function can be expressed in its dual form through the Fenchel-Legendre conjugate as 
\begin{align}
&\left\| \sqrt{\frac{\rho_d}{n}} \Gammam^{1/2}\Bm \ev +\sqrt{\frac{\rho_d}{n}} \Deltam \xv_0-\wv \right\|^2 =\nonumber\\
& \max_{\widetilde\uv}  \widetilde\uv^T \left(\sqrt{\frac{\rho_d}{n}} \Gammam^{1/2}\Bm \ev +\sqrt{\frac{\rho_d}{n}} \Deltam \xv_0-\wv \right)\! -\!\frac{ \| \widetilde\uv \|^2}{4}.
\end{align}
Then, \eqref{eq:Gamma} becomes
\begin{align}\label{}
\Phim^{(n)}: = \min_{\ev } \max_{\widetilde\uv} & \sqrt{\frac{\rho_d}{n}} \widetilde\uv^T \Gammam^{1/2}\Bm \ev +\sqrt{\frac{\rho_d}{n}}  \widetilde\uv^T \Deltam \xv_0 \nonumber\\
&-\widetilde\uv^T\wv -\frac{ \| \widetilde\uv \|^2}{4}+\lambda \rho_d \|\xv_0+\ev \|^2.
\end{align}
One technical requirement of the cGMT is the compactness of the feasibility sets. This can be handled according to the approach in \cite[Appendix A]{thrampoulidis2018precise}, by introducing sufficiently large \emph{artificial} constraint sets $\mathcal{K}_\ev = \{ \ev \in \mathbb{R}^n: \| \ev \|_2 \leq C_e \}$,  and $\mathcal{K}_{\widetilde \uv} = \{ \widetilde \uv \in \mathbb{R}^m: \| \widetilde \uv \|_2 \leq C_{\widetilde{u}} \}$ for some sufficiently large constants (independent of $n$) $C_e, C_{\widetilde{u}} >0$, which will not asymptotically affect the optimisation problem. Then, we obtain
\begin{align}\label{}
\widetilde \Phim^{(n)} = \min_{\ev \in \mathcal{K}_\ev} \max_{\widetilde\uv \in \mathcal{K}_{\widetilde\uv}} & \sqrt{\frac{\rho_d}{n}} \widetilde\uv^T \Gammam^{1/2}\Bm \ev +\sqrt{\frac{\rho_d}{n}}  \widetilde\uv^T \Deltam \xv_0 \nonumber\\
&-\widetilde\uv^T\wv -\frac{ \| \widetilde\uv \|^2}{4}+\lambda \rho_d \|\xv_0+\ev \|^2.
\end{align}
The above optimisation problem is now in the desired min-max form of a PO problem of the cGMT. However, we still have correlated entries in the bi-linear term and we have to transform them to a term that involves a standard Gaussian matrix with i.i.d. entries (as required by the cGMT statement). To do so,
redefine 
\begin{align}
\widetilde\uv = \Gammam^{1/2}\widetilde\uv.
\end{align}
Then, after properly normalising $\widetilde\Phim^{(n)}$ by $\frac{1}{n}$, it becomes
\begin{align}\label{PO_L}
\overline\Phim^{(n)} =& \frac{1}{\sqrt n} \bigg[ \min_{\ev  \in \mathcal{K}_\ev} \max_{\widetilde\uv  \in \mathcal{K}_{\widetilde\uv}}   \frac{\sqrt{\rho_d}}{n} \widetilde\uv^T \Bm \ev +\frac{\sqrt{\rho_d}}{n}   \widetilde\uv^T \Gammam^{-1/2}\Deltam \xv_0 \nonumber\\
&-\frac{1}{\sqrt n}\widetilde\uv^T \Gammam^{-1/2}\wv -\frac{  \widetilde\uv^T \Gammam^{-1} \widetilde\uv}{4\sqrt n}+\frac{\lambda \rho_d }{\sqrt n} \|\xv_0+\ev \|^2 \bigg].
\end{align}
The above optimisation is in a PO form, and its corresponding AO is
\begin{align}\label{AO_1}
\phi^{(n)}:= &\frac{1}{\sqrt n} \bigg[ \min_{\ev \in \mathcal{K}_{\ev}} \max_{\widetilde\uv \in \mathcal{K}_{\widetilde\uv}}  \frac{\sqrt{\rho_d}}{n} \| \ev \| \gv^T \widetilde\uv- \frac{\sqrt{\rho_d}}{n} \| \widetilde\uv \| \sv^T \ev \nonumber\\
&+\frac{\sqrt{\rho_d}}{n} \widetilde\uv^T \Gammam^{-1/2}\Deltam \xv_0-\frac{1}{\sqrt n}\widetilde\uv^T \Gammam^{-1/2}\wv \nonumber\\
&-\frac{  \widetilde\uv^T \Gammam^{-1} \widetilde\uv}{4\sqrt n}+\frac{\lambda \rho_d }{\sqrt n} \|\xv_0+\ev \|^2 \bigg],
\end{align}
where $\gv \in \mathbb{R}^m$ and $\sv \in \mathbb{R}^n$ are independent random vectors with i.i.d. standard normal entries each.

Fixing the norm of the normalised error vector, $\frac{\ev}{\sqrt n}$, to $\eta:= \frac{\| \ev \|}{\sqrt{n}}$, and defining $\bar{\ev}:=\frac{\ev}{\|\ev\|}$ yields
\begin{align}\label{AO_2}
\phi^{(n)}& = \min_{\eta>0} \min_{\| \bar\ev \| =1}\max_{\widetilde\uv \in \mathcal{K}_{\widetilde\uv}}  \frac{\eta \sqrt{\rho_d}}{n}\gv^T \widetilde\uv- \frac{\eta \sqrt{\rho_d}}{n}\| \widetilde\uv \| \sv^T \bar\ev  \nonumber \\
&+\frac{\sqrt{\rho_d}}{n\sqrt n}   \widetilde\uv^T \Gammam^{-1/2}\Deltam \xv_0  -\frac{1}{ n}\widetilde\uv^T \Gammam^{-1/2}\wv-\frac{ 1}{4 n} \widetilde\uv^T \Gammam^{-1} \widetilde\uv \nonumber\\
&+\lambda \rho_d \eta^2+ \frac{2 \lambda \rho_d \eta }{\sqrt n} \xv_0^T \bar\ev + \frac{\lambda \rho_d }{n} \|\xv_0 \|^2.
\end{align}
Defining $\uv :=\frac{\widetilde \uv}{\sqrt n}$ gives:
\begin{align}\label{AO_u}
&\phi^{(n)} = \min_{\eta>0} \max_{\uv \in \mathcal{K}_{\uv}}  \frac{\eta \sqrt{\rho_d}}{\sqrt n}\gv^T \uv+\frac{\sqrt{\rho_d}}{n}  \uv^T \Gammam^{-1/2}\Deltam \xv_0  \nonumber \\
&-\frac{1}{ \sqrt n} \uv^T \Gammam^{-1/2}\wv  -\frac{ 1}{4 } \uv^T \Gammam^{-1} \uv+\lambda \rho_d \eta^2  \nonumber\\
&+ \frac{\lambda \rho_d }{n} \|\xv_0 \|^2\! + \!\min_{\| \bar\ev \| =1}\ \!\frac{\eta }{\sqrt{n}}\left( 2 \lambda \rho_d  \xv_0- \sqrt{\rho_d}\| \uv \|  \sv \right)^T \!\bar\ev,
\end{align}
where $\mathcal{K}_{\uv}$ is defined in a similar fashion to $\mathcal{K}_{\widetilde\uv}$.
The optimisation over $\bar \ev$ can be easily found as follows
\begin{align}\label{eq:w_bar}
 &\min_{\| \bar \ev \| =1} \ \frac{\eta }{\sqrt{n}}\bigg( 2 \lambda \rho_d  \xv_0 - \sqrt{\rho_d}\| \uv \|  \sv \bigg)^T \bar\ev  
 \nonumber\\
 &=-\frac{\eta }{\sqrt{n}} \bigg\| 2\lambda \rho_d \xv_0 - \sqrt{\rho_d}\| \uv \| \sv \bigg \| \nonumber \\
  & \overset{P}{\longrightarrow} -\eta \sqrt{4 \lambda^2 \rho_d^2  + \rho_d \| \uv \|^2},
\end{align}
with a minimiser:
\begin{equation}\label{eq:opt w_bar}
 \widetilde{\bar\ev} = - \frac{2 \lambda \rho_d \xv_0 - \sqrt{\rho_d}\| \uv\|_2 \sv}{\bigg\|  2 \lambda \rho_d \xv_0 - \sqrt{\rho_d}\| \uv\|_2 \sv \bigg\|}.
\end{equation}
Also, note that $\frac{1 }{n} \| \xv_0 \|^2 \pto 1$.
Thus, by applying Lemma 10 in \cite{thrampoulidis2018precise}, we have 
\begin{align}
\widetilde\phi^{(n)} -\phi^{(n)}  \pto 0, 
\end{align}
where
\begin{align}\label{AO_tilde}
\widetilde\phi^{(n)}\! &=\! \min_{\eta>0}  \!  \max_{\uv \in \mathcal{K}_{\uv}} \! \!\frac{1}{\sqrt n}\! \bigg(\!  \eta \sqrt{\rho_d} \gv \!+\!\sqrt{\frac{{\rho_d}}{n}}\! \Gammam^{-1/2}\Deltam \xv_0 \!-\!  \Gammam^{-1/2}\wv \bigg)\!^T \uv \nonumber \\
& -\frac{ 1}{4 } \uv^T \Gammam^{-1} \uv \!+\lambda \rho_d (\eta^2+1)   \!-\eta \sqrt{4 \lambda^2 \rho_d^2  + \rho_d \| \uv \|^2} .
\end{align}
The square root in the last term of the above equation can be written in a variational form using the following identity
\begin{align}
\Theta= \min_{\chi>0} \frac{\chi}{2} +\frac{\Theta^2}{2 \chi},
\end{align}
with $\Theta= \sqrt{4 \lambda^2 \rho_d^2  + \rho_d \| \uv \|^2}.$ Hence, $\widetilde\phi^{(n)} $ becomes
\begin{align}\label{AO_tilde2}
\widetilde\phi^{(n)}& \!= \!\min_{\eta>0} \max_{\substack{ \uv \in \mathcal{K}_\uv \\ \chi>0} } \!\big( \eta \sqrt{\rho_d} \gv\! +\!\sqrt{\frac{{\rho_d}}{n}}  \Gammam^{-1/2}\Deltam \xv_0 \!- \! \Gammam^{-1/2}\wv \big)^T\!\frac{\uv }{\sqrt n}   \nonumber \\
&-\frac{ 1}{4 } \uv^T \Gammam^{-1} \uv\!+\!\lambda \rho_d (\eta^2+1)\! -\! \frac{\eta \chi}{2} \!-\! \frac{\eta (4 \lambda^2 \rho_d^2  + \rho_d \| \uv \|^2)}{2 \chi }.
\end{align}
Next, for convenience, let 
\begin{align}
\widetilde{\gv}: = \eta \sqrt{\rho_d} \gv +\sqrt{\frac{{\rho_d}}{n}}  \Gammam^{-1/2}\Deltam \xv_0 -  \Gammam^{-1/2}\wv,
\end{align}
and 
\begin{align}
\Tm: = \frac{1}{2} \Gammam^{-1} + \frac{\rho_d \eta}{\chi} \Id_m,
\end{align}
then,
\begin{align}\label{}
\widetilde\phi^{(n)} = &\min_{\eta>0} \max_{\substack{ \uv \in \mathcal{K}_{\uv} \\ \chi>0}}  \ \frac{1}{\sqrt n}\widetilde{\gv}^T \uv- \frac{1}{2}\uv^T \Tm \uv \nonumber\\
& -\frac{\eta \chi}{2} - \frac{2 \lambda^2 \rho_d^2 \eta}{\chi} +\lambda \rho_d (\eta^2+1).
\end{align}
The optimisation over $\uv$ is straightforward:
\begin{align}
\uv_* = \frac{1}{\sqrt n}\Tm^{-1} \widetilde{\gv}. 
\end{align}
Then, the AO writes
\begin{align}\label{SOP_1}
 \widetilde\phi^{(n)} = \min_{\eta>0} \max_{ \chi>0} & \ \frac{1}{2n}\widetilde{\gv}^T \Tm^{-1} \widetilde{\gv}^T -\frac{\eta \chi}{2} \nonumber\\
 &- \frac{2 \lambda^2 \rho_d^2 \eta}{\chi} +\lambda \rho_d (\eta^2+1).
\end{align}
The above optimisation is now over scalar variables only, namely $\eta$ and $\chi$ which is easier to analyse. We will refer to \eqref{SOP_1} as the \textit{Scalar Optimisation Problem} (SOP) and study its asymptotic behaviour next.
\subsection{Analysis of the Asymptotic Behaviour of the SOP}
First, note that $\widetilde{\gv}  \sim \mathcal{N}(\mathbf{0}, \Rm_{\widetilde \gv})$, where\footnote{This result is based on the assumption that $\xv_0 = \mathbf{1}_n$.}
\begin{align}
\Rm_{\widetilde \gv} = \rho_d \eta^2 \Id_m + (\rho_d \Rm_{\Delta} +\Id_m) \Gammam^{-1}.
\end{align}
Then, using tools from random matrix theory (RMT) such as the Trace Lemma \cite{Couillet2011}, we have
\begin{align}
 \frac{1}{n}\widetilde{\gv}^T \Tm^{-1} \widetilde{\gv}^T - \frac{1}{n} \tr \left(\Rm_{\widetilde \gv} \Tm^{-1} \right)  \pto 0.
\end{align}
Therefore, again, using \cite[Lemma 10]{thrampoulidis2018precise}, $\widetilde\phi^{(n)} - \overline\phi^{(n)} \pto 0$, where
\begin{align}
\overline\phi^{(n)}:= \min_{\eta>0} \max_{\chi >0}& \frac{1}{2 n} \sum_{j=1}^{m} \frac{\gamma_{j} \rho_d \eta^2 + 1+\rho_d [\Rm_{\Delta}]_{jj}}{\frac{1}{2}+ \frac{\eta \rho_d \gamma_j }{ \chi}}  \nonumber\\
&+ \lambda \rho_d (\eta^2+1) -\frac{\eta \chi}{2} -\frac{2 \lambda^2 \rho_d^2 \eta}{\chi}.
\end{align}
Defining $\mu:=\frac{\chi}{\eta}$, we get 
\begin{align}\label{SOP:eta}
\overline\phi^{(n)}=\min_{\eta>0} \max_{\mu >0}& \frac{1}{2n}\sum_{j=1}^{m} \frac{\gamma_{j} \rho_d\eta^2 + \rho_d [\Rm_{\Delta}]_{jj} +1}{\frac{1}{2} + \frac{ \rho_d \gamma_j }{\mu}} \nonumber \\
&+ \lambda \rho_d(\eta^2+1)  -\frac{\eta^2 \mu}{2}  -\frac{2 \lambda^2 \rho_d^2 }{\mu}.
\end{align}
Finally, note that $\eta$ appears everywhere in $\overline\phi^{(n)}$ as $\eta^2$ and $\eta>0$, so we can use the change of variable $\nu :=\eta^2$ to have
\begin{align}\label{SOP_last}
\overline\phi^{(n)}=\min_{\nu>0} \max_{\mu >0} &\frac{1}{2n}\sum_{j=1}^{m} \frac{\gamma_{j} \rho_d\nu + \rho_d [\Rm_{\Delta}]_{jj} +1}{\frac{1}{2} + \frac{ \rho_d \gamma_j }{\mu}}\nonumber\\
& + \lambda \rho_d(\nu+1) -\frac{\nu \mu}{2}   -\frac{2 \lambda^2 \rho_d^2 }{\mu}.
\end{align}
%
\subsection{Exact Asymptotics of RZF via the cGMT}
We are now in a position to study the asymptotic behaviour of the RZF receiver. 
\subsubsection*{ MSE Analysis}
Let $\widetilde \ev$ be the optimal solution to the AO defined as the solution to \eqref{AO_u}. 
Let $\nu_*$ be the optimal solution to \eqref{SOP_last}. For any $\epsilon>0$, define the set:
\begin{align}\label{MSE:Set}
\mathcal{K}_{\epsilon} = \bigg\{ \pv\in \mathbb{R}^n : \bigg| \frac{1}{n} \| \pv \|^2 - \nu_* \biggr| < \epsilon \bigg\}.
\end{align}
Denote $\hat\eta$ as a minimiser of \eqref{SOP:eta}. By definition, $\hat\eta= \frac{ \| \widetilde\ev \|}{n}$, or using the change of variables that we introduced, $\hat\nu= \frac{ \| \widetilde\ev \|^2}{n}$. We have shown in the previous section that $\phi^{(n)}-\overline\phi^{(n)}\pto 0$, and since $\overline\phi^{(n)}$ in \eqref{SOP_last} has a unique minimiser $\nu_*$, then, applying Lemma 10 in \cite{thrampoulidis2018precise}: $\hat\nu-\nu_*\pto 0$, which implies that
\begin{align}
\bigg| \frac{1}{n}\| \widetilde\ev \|^2 - \nu_*\bigg| \overset{P} \longrightarrow 0.
\end{align}
This proves that $\widetilde \ev \in \mathcal{K}_{\epsilon}$ with probability approaching 1. Then, applying the cGMT yields that  $\widehat\ev \in \mathcal{K}_{\epsilon}$ with probability approaching 1 as well. This ends the proof of Theorem 1.
\subsubsection*{ BER Analysis}
For the BER analysis, we will change the set $\mathcal{K}_{\epsilon}$ in \eqref{MSE:Set} to the set given in \eqref{eq:ber_set}.
\begin{figure*}[h]
\begin{align}\label{eq:ber_set}
\mathcal{K}_{\epsilon}\ =\left\{ \pv \in \mathbb{R}^n : \left| \frac{1}{n} \sum_{i=1}^{n} \!\mathbf{1}_{\{{p}_i \leq -1 \} } \! -Q\left(  \sqrt{\frac{4 \lambda^2 \rho_d (1-\nu_*)+F_{\gamma}(\nu_*,\mu_*)}{\nu_* F_{\gamma}(\nu_*,\!\mu_*)}} \right) \right| <\epsilon\right\}.
\end{align}
\hrule
\end{figure*}

Recall that the optimal solution of the AO in \eqref{eq:opt w_bar} is given as:
\begin{equation}
\widetilde{\ev} = -\hat\eta \sqrt{n} \frac{2 \gamma \xv_0 - \| \uv_*\| \sv}{\bigg\|  2 \gamma \xv_0 - \| \uv_*\| \sv \bigg\|}.
\end{equation}
Also, remember that $\uv_* = \frac{1}{\sqrt n}\Tm^{-1} \widetilde{\gv}$, then, $\| \uv_*\|^2= \frac{1}{n}\widetilde{\gv}^T \Tm^{-2} \widetilde{\gv}^T $.
Then, using the Trace Lemma, we have
\begin{align}
 \frac{1}{n}\widetilde{\gv}^T \Tm^{-2} \widetilde{\gv}^T - \frac{1}{n} \tr \left(\Rm_{\widetilde \gv} \Tm^{-2} \right)  \pto 0.
\end{align}
Or, define
\begin{align}
S_{\gamma}(\nu,\mu) :=& \  \frac{1}{n} \tr \left(\Rm_{\widetilde \gv} \Tm^{-2} \right)\\
&=\frac{1}{n} \sum_{j=1}^{m} \frac{ \rho_d\gamma^2_{j} {\nu} + \rho_d\gamma_j (\rho_d [\Rm_{\Delta}]_{jj} +1)}{\big(\frac{1}{2} + \frac{\rho_d \gamma_j }{\mu}\big)^2} ,
\end{align}
then,
\begin{align}
\| \uv_*\|_2^2- S_{\gamma}(\hat\nu,\hat\mu) \pto0.
\end{align}
Using the fact that $\hat \nu - \nu_* \pto 0$ and $\hat \mu- \mu_* \pto 0$, then for all $i=1,2,\cdots,n$, we have
\begin{align}
\left| \widetilde{e}_i-\frac{-\sqrt{\nu_*} \left(2 \rho_d \lambda - \sqrt{\rho_d S_{\gamma}(\nu_*,\mu_*) } s_i \right)}{\sqrt{4 \rho_d^2 \lambda^2 + \rho_d S_{\gamma}(\nu_*,\mu_*) }} \right| \pto 0.
\end{align}
Hence, using the above expression of $\widetilde \ev$, we have
\begin{align}
&\frac{1}{n} \sum_{i=1}^n {\mathbf{1}}_{\{\widetilde{e}_i \leq -1\}} \nonumber \\
& = \frac{1}{n} \sum_{i=1}^n {\mathbf{1}}_{\{s_i \leq \frac{2\lambda \sqrt{\rho_d \nu_*} -\sqrt{4 \rho_d \lambda^2+S_{\gamma}(\nu_*,\mu_*) }}{\sqrt{\nu_* S_{\gamma}(\nu_*,\mu_*) }}\}},
\end{align}
from which we can easily get
\begin{align}
\left|\frac{1}{n}\!  \sum_{i=1}^n \! {\mathbf{1}}_{\{\widetilde{e}_i \leq -1\}}\! \!
-\! Q\left( \! \! \sqrt{ \! \frac{4 \lambda^2 \rho_d (1-\nu_*) \!+\! S_{\gamma}(\nu_*,\mu_*)}{\nu_* S_{\gamma}(\nu_*,\mu_*)}} \right)\! \right| \!\pto \!0.
\end{align}
Therefore, $\widetilde \ev \in \mathcal{K}_\epsilon$ with probability approaching one.
Note that the indicator function ${\mathbf{1}}_{\{\tilde{e}_i \leq -1\}}$ is not Lipschitz, so we cannot directly apply the cGMT. However, as discussed in \cite[Lemma A.4]{thrampoulidis2018symbol}, this function can be appropriately approximated with Lipschitz functions. Therefore, we can conclude by applying the cGMT that $\widehat{\ev} \in \mathcal{K}_\epsilon$ with probability approaching one, which completes the proof of Theorem~2.
\end{appendices}
\bibliographystyle{IEEEbib}
\bibliography{References.bib}

\begin{IEEEbiography}[{\includegraphics[width=1in,height=1.25in,clip,keepaspectratio]{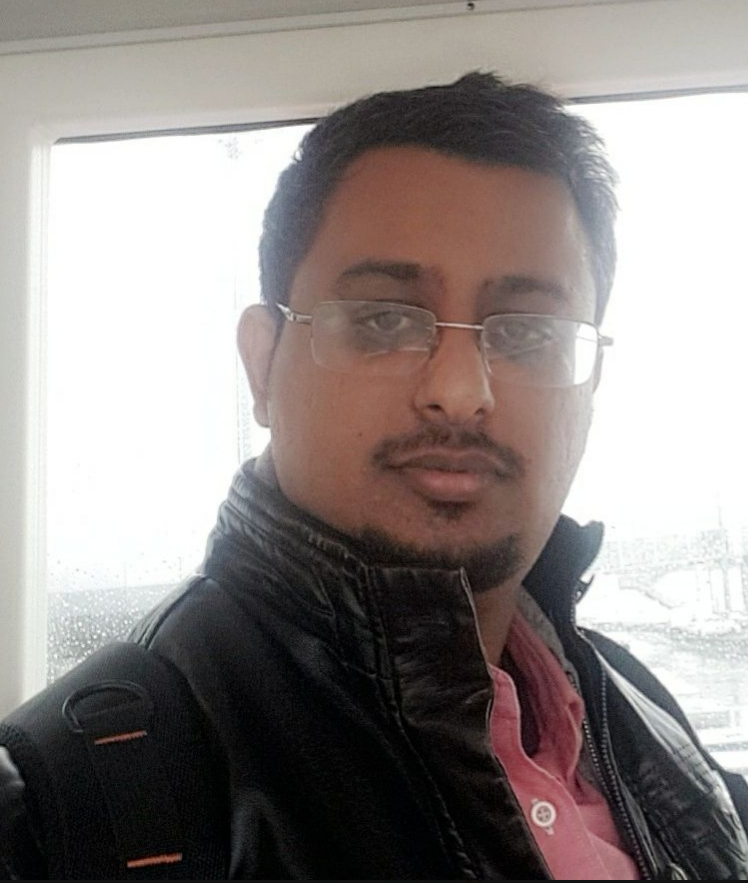}}]{Ayed M. Alrashdi}  received the B.S. degree in Electrical Engineering (with first honors) from University of Ha'il, Ha'il, Saudi Arabia, in 2014, and the M.S. degree in Electrical Engineering from King Abdullah University of Science and Technology (KAUST), Thuwal, Saudi Arabia, in 2016. He received the Ph.D. degree in Electrical and Computer Engineering from KAUST in 2021.

He joined the University of Ha'il in 2014, where he is currently an Assistant Professor in the Electrical Engineering Department.
From 2017 to 2021, he was a Research Assistant with the Information System Lab (ISL) at KAUST. 
His research interests are in the areas of statistical signal processing, high-dimensional statistics, compressed sensing, statistical learning, and wireless communications.
\end{IEEEbiography}
\end{document}